\documentclass[prd,twocolumn,letterpaper,superscriptaddress,floatfix]{revtex4}
\include{epsf}
\usepackage{pslatex}

\begin{document}

\title
{Experimental Limit on the Cosmic Diffuse Ultra-high Energy Neutrino Flux}

\author{P. W. Gorham}
\affiliation{Dept. of Physics \& Astronomy, Univ. of Hawaii at Manoa, 2505 Correa Rd.
Honolulu, HI, 96822}
\affiliation{Jet Propulsion Laboratory, Calif. Inst. of Technology, Pasadena, CA, 91109}
\author{C. L. Hebert}
\affiliation{Dept. of Physics \& Astronomy, Univ. of Hawaii at Manoa, 2505 Correa Rd.
Honolulu, HI, 96822}
\author{K. M. Liewer}
\affiliation{Jet Propulsion Laboratory, Calif. Inst. of Technology, Pasadena, CA, 91109}
\author{C. J. Naudet}
\affiliation{Jet Propulsion Laboratory, Calif. Inst. of Technology, Pasadena, CA, 91109}
\author{D. Saltzberg}
\affiliation{Dept. of Physics \& Astronomy, Univ. of Calif. Los Angeles, Los Angeles, CA}
\author{D. Williams}
\affiliation{Dept. of Physics \& Astronomy, Univ. of Calif. Los Angeles, Los Angeles, CA}

\begin{abstract}
We report results from 120 hours of livetime with the
Goldstone Lunar Ultra-high energy neutrino Experiment (GLUE).
The experiment searches for $\leq 10$ ns microwave pulses from
the lunar regolith, appearing in coincidence at two large
radio telescopes separated by 22~km and linked by
optical fiber. Such pulses would arise from subsurface 
electromagnetic cascades induced by interactions of
$\geq 100$ EeV neutrinos in the lunar regolith. 
No candidates are yet seen, and the implied limits
constrain several current models for ultra-high energy neutrino
fluxes. 
\end{abstract}

\maketitle


In 1962, G. Askaryan predicted 
that electromagnetic cascades in dense media should produce strong coherent
pulses of microwave Cherenkov radiation~\cite{Ask62}.
Recent confirmation of this hypothesis at
accelerators~\cite{Sal01} strengthens the 
motivation to search for such emission from cascades
induced by predicted high energy neutrino fluxes,
closely related to the measured fluence of $\simeq 10^{20}$ eV
cosmic rays in many models. 

Two such models, the
Z-burst model~\cite{Wei99}, and a generic class 
known as Topological Defect (TD) models~\cite{Yos97}, predict ultra-high
energy (UHE) neutrinos with either
monoenergetic or very hard energy spectra. In the
Z-burst model, UHE neutrinos annihilate
with relic cosmic background neutrinos
via the $\nu\bar{\nu} \rightarrow Z_0$ channel. The $Z_0$
then decays rapidly in a burst of hadronic secondaries
which create the observed $\sim 10^{20}$~eV cosmic rays.
The need to match the observed UHE cosmic ray
fluxes and satisfy the current constraints on neutrino masses
(which modify the annihilation resonance energy) then
lead to a requirement on minimal neutrino fluxes at the
resonance energy near $10^{22-23}$~eV. The Z-burst model
thus formally requires only neutrinos at a single energy,
with no specification for how such a flux might be
produced.

The Z-burst model is also significant in that it is a
variation on an earlier idea~\cite{Weiler82} in which
the $\nu\bar{\nu}$ annihiliation process could be used
as a probe of the cosmic background neutrinos, one of the
few viable ways ever proposed for detection of these 
relic cosmological neutrinos--it requires only a sufficient
flux of UHE neutrinos and a detector with the
sensitivity to measure them. 
Constraints on these UHE $\nu$ fluxes thus can rule out
this potential detection channel for the relics, in addition
to excluding their role in UHE cosmic ray production.

TD models, in contrast, postulate a very massive relic particle
from the early universe which is decaying in the current epoch and
producing secondaries observed as UHE cosmic rays. The
required masses approach the Grand-Unified Theory (GUT) scale
at $\sim 10^{24}$~eV, and the decay products thus have a very hard
spectrum extending up to the rest mass energy of the particles.
Because of these very hard spectra, detectors optimized for
lower-energy neutrinos, even up to PeV energies, do not
yet constrain these models, and new approaches, such as 
the experiment we report on here, are required.

Neutrinos with energies above 100 EeV (1 EeV = $10^{18}$~eV)  
can produce cascades in the upper 10~m of the 
lunar regolith resulting in pulses that are detectable
at Earth by large radio telescopes.~\cite{Zhe88,Dag89}  One
prior experiment has been reported, using the Parkes
64 m telescope~\cite{H96} with 10 hours of livetime. 
In the decimeter band, the signal should appear
as highly linearly-polarized, band-limited 
electromagnetic impulses~\cite{zhsa}.
However, since there are many anthropogenic sources of 
impulsive radio-frequency interference (RFI), the primary problem in
detecting neutrinos is to reject such interference. 
Since 1999 we have conducted a series of experiments 
in search of such pulses, using the
JPL/NASA Deep Space Network antennas at Goldstone, CA~\cite{Gor99}.
We have essentially eliminated RFI background by
employing two antennas in coincidence.


Although the total livetime 
accumulated in our experiment is a relatively small fraction
of what is possible with a dedicated system, the volume of 
material to which we are sensitive is enormous, exceeding
100,000 km$^3$ at the highest energies. The resulting 
sensitivity is enough to begin constraining some
models for diffuse neutrino fluxes at energies above $10^{20}$ eV.
We report here on results from 120 hours of livetime.

	
The lunar regolith is an aggregate layer of fine particles and
small rocks, consisting mostly of silicates and
related minerals, with meteoritic
iron and titanium compounds at an average level of several per cent, and
traces of meteoritic carbon. Its depths are 10 to 20~m in the maria and valleys, 
but may be hundreds of meters in portions of the highlands~\cite{Mor87}.
It has a mean dielectric constant
of $\epsilon=n^2\simeq 3$ and a density of $\rho \simeq 1.7$ g cm$^{-3}$,
both increasing slowly with depth. Measured values for the loss tangent 
vary widely depending on iron and titanium content, but a
mean value at high frequencies is $\tan \delta \simeq 0.003$,
implying a field attenuation length  $L_{\alpha}\simeq~9$~m at 2~GHz~\cite{Olh75}.

\begin{figure} 
\epsfxsize=3.2in
\epsfbox{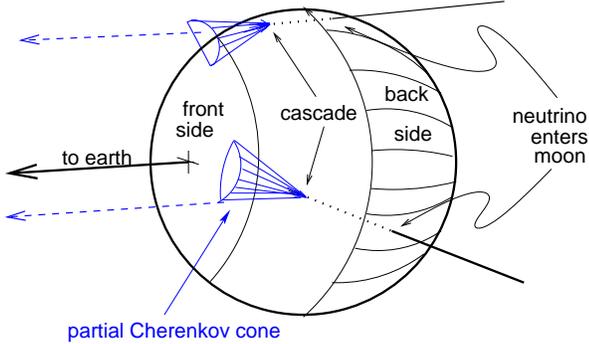}
\vspace{10pt}
\caption{Geometry of lunar neutrino cascade event detection.}
\label{moongeom}
\end{figure}

Fig.~\ref{moongeom} illustrates the signal emission 
geometry. At 100 EeV the interaction length of a
neutrino 
is about 60~km~\cite{Gan00}.
Upon interaction, a cascade $\sim 10$~m long
forms, and Compton scattering, positron annihilation, and other
processes lead to a $\sim 20\%$ negative charge excess.
This cascade radiates a cone of coherent Cherenkov emission 
at an angle from the shower axis of 
$\theta_C = \cos^{-1}(1/\beta\sqrt{\epsilon}) \simeq 54^{\circ}$, 
with an angular spread
of $\Delta \theta \simeq 1^{\circ}$ at 2~GHz.
At an ideal smooth regolith surface, the refraction obeys Snell's
law, and the exit angle $\theta'$ is magnified by the gradient
$\Delta \theta'/\Delta \theta = n\cos\theta / \sqrt{1-n^2\sin^2\theta}$
which equals $n$ for normal incidence ($\theta=0$), but becomes much larger
as $\theta$ approaches the total-internal reflectance angle
$\theta_{TIR} = \pi/2 - \theta_C$. This magnification improves
the acceptance for an earth-based detector, at the expense
of an increase in energy threshold. Our ray tracing shows that similar effects
obtain statistically for a more realistic surface as well.


For our search we used the shaped-Cassegrainian
70~m antenna DSS14, and the 34~m beam-waveguide antenna DSS13,
separated by 22~km.
The S-band (2.2~GHz)
right-circular-polarization (RCP) signal from DSS13 is 
filtered to 150~MHz bandwidth and down-converted to an
intermediate frequency (IF) near 300~MHz.
The band is subdivided into high and low frequency
halves with no overlap. 
The DSS14 dual polarization S-band signals are down-converted
to the same 300~MHz IF, and a combination of bandwidths from
40-150~MHz are used for sub-band triggering on impulsive signals.
At DSS14, an  L-band (1.8 GHz) feed which is off-pointed by $\sim 0.5^{\circ}$
produces a 40~MHz bandwith monitor of terrestrial interference signals.



\begin{figure} 
\epsfxsize=3.2in
\epsfbox{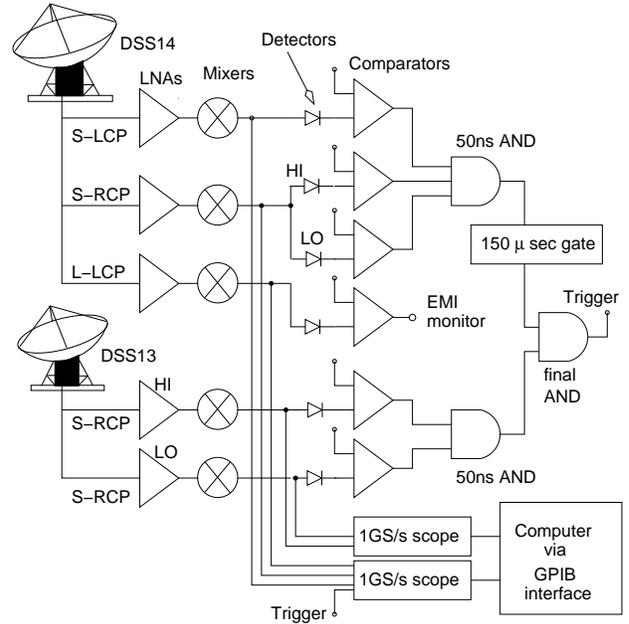}
\vspace{10pt}
\caption{GLUE trigger and data recording system.}
\label{trigger1}
\end{figure}

Fig.~\ref{trigger1} shows the layout of the trigger. The 
signals from the two antennas are converted to
unipolar pulses using tunnel-diode detectors with 
a $\sim 10$~ns integration time. A comparator then
test for pulses above threshold, and
a local coincidence within 50~ns is 
formed among the channels at each antenna.
The DSS14 coincidence 
between both circular polarizations ensures that the signals
are highly linearly polarized, and the
split-channel coincidences
ensure that the signal is broadband.

A global trigger is formed between the local coincidences of the
two antennas within  a 150~$\mu$s window, which 
encompasses the possible geometric delay range for
the Moon throughout the year.  Although use of a smaller window is possible,
a tighter coincidence is applied offline and the out-of-time events
provide a large background sample.
Upon the global coincidence, 
a 250~$\mu$s record, sampled at
1~Gsamples/s, is stored. The average trigger rate, due primarily
to random coincidences of thermal noise fluctuations, 
is $3\times 10^{-3}$~Hz. Terrestrial
interference triggers are on average a few percent of the total, but
we have averaged $\geq 95$\% livetime during the runs to date.



  The precise geometry of the experiment is a crucial
discriminator for events from the Moon. 
The relative delay between
the two antennas is $\tau = c^{-1} |\vec{B}| \cos \theta$ where
$\theta$ is the apparent angle of the Moon with respect to the
baseline vector, $\vec{B}$.
For our 22~km baseline, we have a
maximum delay difference of $\tau_{max} = \pm 73~\mu$s. 
Detectable events can occur anywhere on the Moon's surface within the
antenna beam.
This produces
a possible spread of 630~ns in the differential delay of the received pulses
at the two antennas.

The 2.2~GHz antenna beamwidths 
between the measured first Airy nulls are
$0.27^{\circ}$ for the 70~m, and $0.56^{\circ}$ for the 34~m.
We took data in three configurations: pointing at the limb, 
the center, and halfway between. The measured source temperatures
varied from 70K at the limb, to 160K at the Moon center, with
system temperatures of 30-40K.


Timing and amplitude calibration are accomplished in several steps.
We internally calibrate the
back-end trigger system using a synthesized IF pulse signal,
giving precision of order 1~ns.  We use a
pulse transmitter (single-cycle at 2.2GHz) aimed at the antennas 
to calibrate the cross-channel delays of each antenna
to a precision of 1~ns. The cross-polarization 
timing at DSS 14 is also checked 
since the thermal radiation from the limb of the Moon
is significantly linearly polarized (from differential Fresnel
effects~\cite{Soboleva_63, Heiles_63}),
introducing an easily detectable LCP-to-RCP correlation.

Dual antenna timing calibration is accomplished by cross-correlating
a 250 $\mu$s thermal noise sample of a bright quasar, typically
3C273, recorded from both antennas at the same time and in the
same polarization, using the 
identical data acquisition system used for the pulse detection.
This procedure establishes the global $\sim$136~$\mu$s
delay between the two antennas to better than 10~ns.
Amplitude calibration is accomplished by 
referencing to a known thermal noise source.
The system temperature during a run
fixes the value of the noise level and therefore the 
energy threshold.  We also check the system linearity
using pulse generators to ensure that the entire system 
has the dynamic range to see large pulses. 

\begin{figure}
\epsfxsize=3.2in
\epsfbox{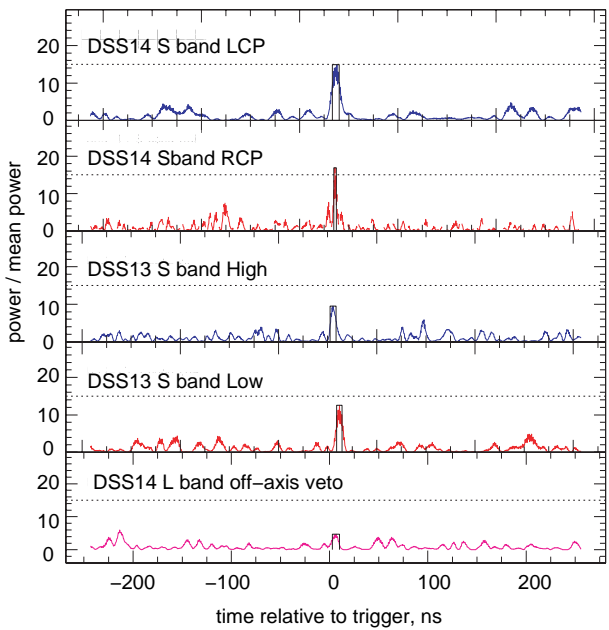}
\vspace{10pt}
\caption{An example of a GLUE triggered event.
}
\label{event}
\end{figure}

Figure~\ref{event} shows a typical event which triggered the
system. The top two panes contain
the DSS14 LCP and RCP signals, and a narrow pulse is present in both
polarizations, indicating a broadband spectral content, and
a high degree of linear polarization. 
The pulse power is normalized to the local mean power over a
250 $\mu$s window. In the
third and fourth panes, the two channels from DSS13 are shown. 
In the fifth pane from the top  
the L-band offset feed signal from DSS 14 is shown, and 
no RFI is present.  The measured delay relative to
the expected Moon time is $-1.1~\mu$s in this case, slightly
larger than allowed. Systematic timing offsets
from channel to channel are well under 10~ns.

Two largely independent analyses look for pulses 
corresponding to an electric field 
$6\sigma$ above thermal noise in all channels.
Both analyses remove terrestrial RFI with either a visual or algorithmic
method.   Each enforces
$\sim 20$~ns local coincidence timing at each antenna.
The precise values of the cuts were determined and fixed
before looking at more than half of the data.
We have seen no candidates in 120 hours of livetime.  
The signal efficiency of the RFI and timing cuts is estimated to be
$>98$\%. To estimate background levels, 
we also search of order 100 different delay values that are 
inconsistent with the Moon's sky position, and we have also found no
candidates in this search. Hence we observe no events with a
background of $\leq 0.01$ events.

The field strength (V/m/MHz)
from a cascade of total energy $W_T$ in regolith material
can be expressed~\cite{zhsa}:
\begin{equation}
\label{zhs}
E = \frac{1.45 \times 10^{-7}}{R}
\ \frac{W_{T}}{1 {\rm TeV}} \ \frac{\nu}{\nu_0}
\ \frac{1}{(1 + (\nu/\nu_0)^{1.44})}
\end{equation}
where  $R$ is the distance to the source in meters, 
$\nu$ is the radio frequency, and 
the decoherence frequency is $\nu_0 \simeq 2500$ MHz for the regolith
($\nu_0$ scales mainly by radiation length). For typical
parameters in our experiment, a $10^{20}$ eV cascade will
result in a peak field strength at Earth of $E \simeq 1.3~\mu$V m$^{-1}$
for a 70 MHz bandwidth.
We have verified Equation~\ref{zhs} to within a factor of $2$
through accelerator tests~\cite{Sal01} using silica sand targets
and $\gamma$-ray-induced cascades with composite $W_T \sim 10^{19}$~eV.




Based on the effective antenna aperture $\eta A$
and the fact that the background
events are due to fluctuations in the black-body power of the Moon,
we estimate that the 
minimum detectable field strength for a linearly polarized pulse  is
\begin{equation}
E_{min} ~=~ N_{\sigma} \sqrt{{2 k T_{sys} Z_0 \over  \eta A \Delta \nu \sqrt{\epsilon}  }},
\end{equation}
where $Z_0=377 \Omega$, and
$N_{\sigma}$ the number of standard deviations required per channel 
relative to thermal fluctuations.
For the lunar observations on the limb, which make up about 85\% of
the data reported here, $T_{sys} \simeq 110$ K (including the source
contribution), $\nu = 2.2$~GHz,
and the average $\Delta \nu \simeq 70$ MHz. For the 70 m antenna,
with $\eta \simeq 0.8$, the minimum detectable
field strength at $N_{\sigma}=6$ is $E_{min} \simeq 0.8\mu$V~m$^{-1}$. 
The estimated cascade threshold energy for
these parameters is $W_{thr} = 6 \times 10^{19}$~eV. Since
the mean inelasticity is $\langle y \rangle = 0.2$,
the threshold neutrino energy is $\sim 3 \times 10^{20}$~eV.


Effective volume and acceptance vs. neutrino energy has been
estimated via two independent Monte Carlo simulations, including
the current estimates of both charged and neutral current
cross sections~\cite{Gan00}, and the $y$ distribution.
The neutrino species are assumed to be fully mixed upon arrival.
All neutrino flavors were included,
and LPM effects in the shower formation
were estimated~\cite{zhsa}. At each
neutrino energy, a distribution of cascade angles and depths with respect
to the local surface was obtained, and a refraction propagation
of the predicted Cherenkov angular distribution
was made through the regolith surface, including absorption, reflection,
and roughness effects, and thermal noise fluctuations in the detector.




Via ray-tracing in our simulations (see Ref.~\cite{Gor01}), we
find that, although the specific flux density 
of the events are lowered by a factor of order 2 from 
refraction and scattering, the effective volume and acceptance
solid angle are increased by as much as an order of magnitude. The 
neutrino acceptance solid angle
is thus about a factor of 20 larger than the apparent solid angle of the 
Moon itself. UHE cosmic rays could produce a background
for lunar neutrino detection; however, there are several processes that 
suppress cosmic ray radio emission with respect to that of neutrinos,
including total internal reflection, and formation zone effects~\cite{Gor01}.
The level of the potential UHE cosmic ray background is unknown,
but as we have detected no events of any kind, our limits obtain in any case.

\begin{figure} %
\epsfxsize=3.9in
\epsfbox{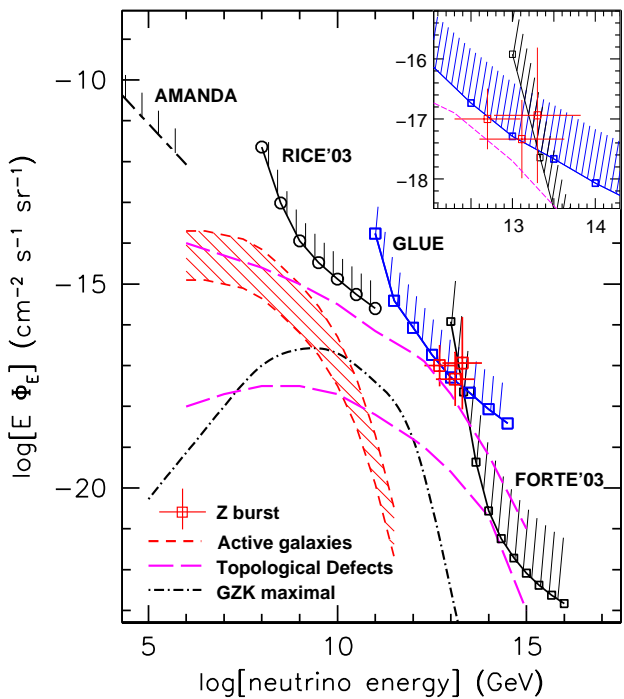}
\vspace{10pt}
\caption{Model neutrino fluxes \& limits from the present work
and other current experiments. The inset expands the region near
the Zburst predictions. }
\label{limits}
\end{figure}

Figure~\ref{limits} plots the predicted fluxes of 
ultra-high energy (UHE) neutrinos from AGN production~\cite{Man96,Steck91},
a maximal flux from UHE cosmic-ray interactions~\cite{Engel01},
two topological defect models~\cite{Yos97}, and
the Z-burst scenario~\cite{Wei99}. Two other current
limits in this energy regime are plotted, 
from the RICE experiment~\cite{RICE03}, and the FORTE 
satellite~\cite{FORTE03}.
Our 90\% confidence level, differential model-independent 
limit for 120 hours of livetime
is shown plotted with large squares~\cite{numbers}, based
on the observation of no events above an equivalent $6\sigma$ level
amplitude (referenced to the 70 m antenna).

All of the {\em differential} limit curves in Fig.~\ref{limits} 
(except for the $E_{\nu_{\mu}}^{-2}$ power law AMANDA limit) 
correspond to the inverse of the
energy-dependent exposure = (neutrino aperture $\times$ time)~\cite{Mlimits} for each
detector noted, scaled by the Poisson upper-limit factor
(2.3 for 90\% CL). Our plotted model-independent limit curves are 
very conservative, and
become more restrictive when integrals over specific 
broad-spectrum models are considered.
When converted to
{\em integral} form, our limits constrain power-law neutrino models that
parallel our curves for $\sim 1$ decade, even a
factor of 3-5 below our plotted differential limits.

For quasi-monoenergetic models such as the Z-burst scenarios, 
the limits are exact, and the combined GLUE and FORTE results 
constrain much of the available Z-burst parameter space.
Models such as the highest TD curve
shown~\cite{Yos97} are largely excluded at higher 
energies by the GLUE and FORTE results,
since the {\em integral} (flux$~\times~$aperture)
would lead to $\sim 3$ and $\sim 10$
events in each detector, respectively, in conflict with the null results.

We thank M. Klein, T. Kuiper, R. Milincic, and the 
Goldstone staff for their enthusiastic support.
We are especially grateful to George Resch (d.2001)
whose encouragement and unflagging support made this
work possible. The work was performed in part
at the Jet Propulsion Laboratory, California Institute of Technology, under 
contract with NASA, the Caltech President's
Fund, DOE contracts at Univ. of Hawaii and UCLA, 
the Sloan Foundation,
and the National Science Foundation.

 

\begin{thebibliography}{99}
\bibitem{numbers} The values of the 8 plotted points are:
-13.8,-15.4,-16.1,-16.7,-17.3,-17.7,-18.1,-18.4.
\end{thebibliography}
\end{document}